\documentclass[prb,aps,epsf,twocolumn,showpacs,10pt,superscriptaddress,floatfix
]{revtex4-1}
\usepackage{graphicx,amsfonts,times,bm,amsmath,verbatim,color,array}

\newcommand{\bra}[1]{\langle {#1} |}
\newcommand{\ket}[1]{| {#1} \rangle}

\newcommand{\e}{\varepsilon}

\usepackage{natbib}
\usepackage{graphicx,amssymb,amsmath,ifthen,braket,xcolor}
\usepackage{bm}

\newcommand{\angstrom}{\mbox{\normalfont\AA}}

\begin{document}


\title{Non-linear Nernst effect in bilayer WTe$_2$} 



\author{Chuanchang Zeng}
\affiliation{Department of Physics and Astronomy, Clemson University, Clemson, SC 29634, USA}
\author{Snehasish Nandy}
\affiliation{Department of Physics, Indian Institute of Technology Kharagpur, West Bengal 721302, India}
\affiliation{Department of Physics, University of Virginia, Charlottesville, VA 22904, USA}
\author{A. Taraphder}
\affiliation{Department of Physics, Indian Institute of Technology Kharagpur, West Bengal 721302, India}
\affiliation{School of Basic Sciences, Indian Institute of Technology Mandi, Kamand 175005, India}
\author{Sumanta Tewari}
\affiliation{Department of Physics and Astronomy, Clemson University, Clemson, SC 29634, USA}


\begin{abstract}
Unlike the linear anomalous Nernst effect, the non-linear anomalous Nernst effect (NLANE) can survive in an inversion symmetry broken system even in the presence of time-reversal symmetry. Using semiclassical Boltzmann transport theory, we study the non-linear anomalous Nernst effect that arises as the second-order response function to the applied temperature gradient. We find that the non-linear Nernst current, which flows perpendicular to the temperature gradient even in the absence of a magnetic field, arises due to the Berry curvature of the states near the Fermi surface, and thus is associated with purely a Fermi surface contribution. We apply these results to bilayer WTe$_2$, which is an inversion broken but time reversal symmetric type-II Weyl semimetal supporting chiral Weyl fermions. By tuning the spin-orbit coupling, we show that the sign of the NLANE can change in this system. Together with the angular dependence, we calculate the temperature and chemical potential dependencies of NLANE in bilayer WTe$_2$, and predict specific experimental signatures that can be checked in experiments.
\end{abstract}


\maketitle

\section{Introduction} \label{sec:i}
The Berry phase effects in anomalous transport phenomena driven by the gradient of temperature or chemical potential is now well-developed theoretically and have been seen in experiments  \cite{Dxiao2006_thermolelectric,Dxiao2010_berryphase,Nagaosa2010_AHE,Tokura2001_AHE,Fang2003_AHE,Yu2010_AHE,qikun2013_AHE}. A nonzero Berry curvature along with the nonuniform statistical distribution of carriers resulting from the external fields have been  used  to explain the transport anomalies like the anomalous Hall effect (AHE)\cite{Haldane2004_AHE,qikun2013_AHE,Yu2010_AHE,Niu2002_AHE,Lee2004_AHE,Fang2003_AHE,Tokura2001_AHE,Nagaosa2003_AHE,Xu2011_AHE,Qiao2010_AHE} and anomalous Nernst effect (ANE) \cite{Xu2000_ANE,yyWang2001_ANE, Lee2004_ANE, yyWang2006_ANE,cZhang2008_ANE,cZhang2009_ANE,zgZhu2013_ANE, Girish2016_ANE,Lundgren2014_ANE,kim2014_ANE,Girish2018_ANE,Saha2018_ANE}. These Berry curvature-related topological transport anomalies are linear responses, which require the breaking of time reversal (TR) symmetry  by  a complex order parameter or internal magnetization \cite{qikun2013_AHE,cxliu_2008_breakingTRS,Nagaosa_2016_breakingTRS,Burkov2014_breakingTRS,Niu2002_AHE,Yu2010_AHE,Xu2011_AHE}. With the increasing interest in non-linear properties of topological materials \cite{Moore2010_NL,Steve2012_NL,Nagaosa2016_NL,Alexander2013_NL,Zhang2008_NL,Wu2016_NL,Chen2014_NL,JMoore2016_NL}, the second-order non-linear Hall effect was proposed as a new type of Hall effect that could survive even in the TR invariant systems with broken inversion symmetry\cite{Fu2015_NLAHE,JMoore2016_NL}. Rather than being induced by the Berry curvature itself, the non-linear response  depends  on the Berry curvature dipole (BCD), a moment of the Berry curvature over the occupied states\cite{Fu2015_NLAHE, yZhang2018_BCD, Facio2018_BCD}.

Monolayer transition metal dichalcogenides (TMDCs) have been proposed as the platforms supporting the non-linear Hall effect due to their large spin-orbit coupling and non-centrosymmetric band structure \cite{Dxiao2012_TMDCs,gbLiu2013_TMDCs,xiaodong2014_TMDCs}. Recently, it has been reported that the non-linear Hall currents can occur as a second-order response to the applied external electric field in TR invariant but inversion symmetry breaking materials, especially the TMDCs and Weyl semimetals (WSMs) \cite{Zhang2018_NLAHE,qMa2019_NLAHE,kKang2019_NLAHE,Law2019_NLAHE}. The underlying physics associated with the non-linear anomalous Hall effect (NLAHE) has been shown to be related to the band structure properties. Several candidate materials with low crystalline symmetries have been proposed to exhibit a finite or enhanced BCD \cite{zzDu2019_NLAHE_1,zzDu2019_NLAHE_2,Niu2019_NLAHE,Low2019_BCD,Nandy2019_NLAHE}.

WTe$_2$ as a type-II WSM exhibiting chiral Weyl Fermions that break Lorentz invariance\cite{Bernevig2015_WTE2,Shi2019_WTE2}, has attracted tremendous attention in the studies of transport anomalies \cite{Qian2014_QSHE,Fei2017_QSHE,Wu2018_QSHE, Seifert2019_phoInduced,Xu2018_switchable}. The topological properties of WTe$_2$ are caused by its Berry curvature concentrated around the Weyl points with linear band-crossing, acting like monopoles of Berry curvature whose value for the valence band and conduction band are opposite \cite{wan2011_WTE2}. The nontrivial phenomena in TR invariant systems such as WTe$_2$ requires an asymmetric distribution of the Fermi occupations at $\bm{k}$ and $\bm{-k}$. Interestingly, a few-layered WTe$_2$ system, e.g. a bilayer WTe$_2$, intrinsically breaks the inversion symmetry and leaves a nonzero Berry curvature distribution on the Fermi surface. It is the rearrangement of the positive and negative Berry curvature in different momentum that leads to the so-called BCD. Specifically, due to the big net BCD on a given Fermi plane, the bilayer WTe$_2$ system shows a strong non-linear Hall response when a small spin-orbit torque is induced \cite{zzDu2019_NLAHE_1,zzDu2019_NLAHE_2,Law2019_NLAHE}. The strong tunability of bilayer WTe$_2$ by electric gating and/or strain makes the discovery of non-linear phenomena in inversion-symmetry-breaking 2D materials particularly promising.

\begin{figure}[h]
	\begin{center}
		\includegraphics[width=0.47\textwidth]{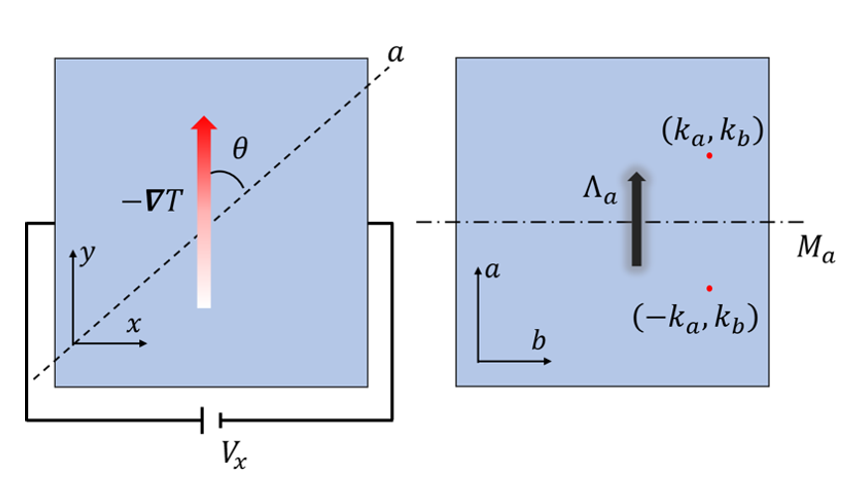}
		\llap{\parbox[b]{155mm}{\large\textbf{(a)}\\\rule{0ex}{38mm}}}
		\llap{\parbox[b]{75mm}{\large\textbf{(b)}\\\rule{0ex}{38mm}}}
	\end{center}
	\caption{(Color online) (a) Schematic experimental setup for measuring the non-linear anomalous Nernst effect in a time-reversal symmetric but inversion symmetry broken system. $V_x$ is measured as the non-linear anomalous Nernst voltage when a temperature gradient ($-\bm{\nabla} T$, represented by the red arrow) is applied along $y-$direction forming an angle $\theta$ with the principal axis $a$ (black dashed line). (b) $M_a$ is the mirror line which takes $k_a$ to $-k_a$. Due to the mirror symmetry and time reversal symmetry, only the $a-$component (black arrow) of the non-linear anomalous Nernst coefficient ($\bm{\Lambda}$), which is perpendicular to the mirror line $M_a$, is non-zero.   	\label{fig:schemati}}
\end{figure}

In the linear regime, the ANE originating from the Berry curvature describes the generation of a charge current in the presence of a transverse temperature gradient and a broken TR symmetry, and vanishes when the TR symmetry is preserved. However, as a counterpart of the NLAHE revealed by charge current, the non-linear anomalous Nernst effect could possibly be non-zero even in the presence of TR symmetry and measured through the heat transport.
Recently the intrinsic non-linear anomalous thermoelectric effects have been proposed for the loop-current model of the cuprate superconductors where the combined TR symmetry and inversion symmetry is retained \cite{Gao2018_NLANE}. However, in this model, the TR symmetry is still broken individually due to the orbital toroidal moment. Very recently, a Hamiltonian of TMDCs under uniaxial strian has been used to demonstrate the NLANE within the two-dimensional TR invariant system with broken inversion symmetry\cite{xiaoqin2019_NLANE}. Here the non-linear anomalous Nernst current is found to have a different temperature dependence in the high  and low temperature regimes.
In another recent work on non-centro-symmetric crystals\cite{Naoto_2019}, the non-linear thermo-electric conductivity has been derived up to the second order to the applied thermal gradient by considering a thermally induced nonlinear perturbation to the Fermi distribution function. In contrast to these works, in this paper we focus on bilayer WTe$_2$, because this system has recently been successfully used to experimentally demonstrate the NLAHE in a TR invariant system\cite{qMa2019_NLAHE,kKang2019_NLAHE}. Because of the existence of nonzero NLAHE, we also expect a nonzero NLANE in this system from Onsager reciprocity. Our calculations and results in this paper will thus be of immediate experimental relevance for the demonstration of NLANE in a system that has already been shown to support NLAHE. In the recent experiments of the NLAHE in bilayer WTe$_2$\cite{kKang2019_NLAHE,qMa2019_NLAHE}, a transverse voltage drop at the second-harmonic frequencies is found quadratically dependent to the longitudinal driving current. In analogy with the NLAHE, a second-harmonic type response is also possible in principle for the NLANE provided the thermal gradient is time dependent. In what follows, we perform a systematic derivation of the NLANE in a TR invariant but inversion symmetry broken system based on the Boltzmann semiclassical approach. We then provide the general angular dependence of the NLANE response relevant for the Nernst experimental setup (Fig.~(\ref{fig:schemati})). We apply these results to the case of bilayer WTe$_2$ and make several experimental predictions, including the dependence of the NLANE conductivity on various experimental parameters such as temperature ($T$), chemical potential ($\mu=E_F$), inter-layer coupling ($\gamma$), tilting of the Dirac cones ($t_x$), and spin orbit coupling ($\eta$).

This paper is organized as follows: In Sec. \ref{sec:ii} we discuss the Boltzmann semiclassical approach to systematically calculate the NLANE response. We derive the  general expressions for the NLANE in the presence of a temperature gradient in an appropriate Nernst setup. In Sec. \ref{sec:iii}, we develop the general angular dependence of the NLANE response, where the angle refers to that between the principal axes and applied temperature gradiant (Fig.~\ref{fig:schemati}(a)). In Sec.~\ref{sec:3ii}, we apply these general semiclassical results to the specific case of bilayer WTe$_2$ described in Sec.~\ref{sec:3i} and make several experimental predictions. We end with a conclusion in Sec. \ref{sec:3iii}.

\section{Semiclassical Boltzmann formalism of Non-linear Nernst response} \label{sec:ii}

In addition to the band energy, the Berry curvature of the Bloch bands is required for a complete description of the electron dynamics in topological systems. Therefore, the transport properties get substantially modified due to the presence of non-trivial Berry curvature of the Bloch bands\cite{Dxiao2010_berryphase,Dxiao2006_thermolelectric}. The Berry curvature of the $i^{th}$ band for a Bloch Hamiltonian is defined as
\begin{equation}
    \begin{split}
     \Omega^{a^{\prime}}_{i\bm{k}}=-2\epsilon_{a^{\prime}b^{\prime}c^{\prime}} \sum_{j\neq i}\frac{\operatorname{Im}\left(\bra{i} \partial_{k_{b^{\prime}}} H\ket{j}\bra{j} \partial_{k_{c^{\prime}}} H \ket{i}\right)} {\left(\e^i_{\bm{k}}-\e^j_{\bm{k}}\right)^2}
    \end{split}  \label{eq:berrycur}
\end{equation}
where $\ket{i}$ is the eigenvector for the $i^{th}$ band with eigenenergy $\e^i_{\bm{k}}$ and $\epsilon_{a^{\prime}b^{\prime}c^{\prime}}$ is the Levi-Civita symbol.
The general form of the Berry curvature can be obtained using symmetry analysis. From the Eq.~(\ref{eq:berrycur}), it is clear that the Berry curvature follows $\mathbf{\Omega_{-k}}=-\mathbf{\Omega_{k}}$ under TR symmetry. On the other hand if the system has spatial inversion symmetry, then it follows $\mathbf{\Omega_{-k}}=\mathbf{\Omega_{k}}$. Therefore, for a system with both TR and spatial inversion symmetry the Berry curvature vanishes identically throughout the Brillouin zone~\cite{Dxiao2010_berryphase}. However, in the presence of broken inversion and/or TR symmetry, the Berry curvature of the Bloch bands can be non-trivial.

In the presence of non-zero Berry curvature, the conventional semiclassical equation of motion for an electron becomes modified by adding a transverse anomalous term to the velocity, given by\cite{Dxiao2006_thermolelectric,Dxiao2010_berryphase,Sundaram1999_berryphase},
\begin{equation}
\begin{split}
\bm{\dot{r}} &=\frac{1}{\hbar}\frac{\partial  \e_{ \bm{k}}}{\partial \bm{k}} +\frac{\bm{\dot{p}}}{\hbar} \bm{\times} \bm{\Omega_k}  \\
\bm{\dot{p}} &=e \bm{E} + e \bm{\dot{r} \times B},
\end{split}   \label{eq:semic_eom}
\end{equation}
where $\e_{\bm{k}}$ is the energy dispersion for given momentum $\bm{k}$, and $\bm{E}, \bm{B}$ are the external electric and magnetic fields respectively.
The above coupled equations for $\bm{\dot{r}}$ and $\bm{\dot{p}}$ could be solved together to get\cite{Duval2006_solvetogether,Son2012_solvetogetheR,JMoore2016_NL},
\begin{equation}
\begin{split}
\bm{\dot{r}} &= D\left( \bm{B},\bm{\Omega_k}\right)
\left[ \bm{v_k}+\frac{e}{\hbar} \left(\bm{v_k}\cdot \bm{\Omega_k}\right) \bm{B} +  \frac{e}{\hbar} \left( \bm{E \times \Omega_k}\right)\right]  \\
\bm{\dot{p}} &=D\left( \bm{B},\bm{\Omega_k}\right) \left[ \frac{e}{\hbar} \bm{v_k}\bm{\times} \bm{B} +e\bm{E} +\frac{e^2}{\hbar} \left( \bm{E \cdot B}\right)\bm{\Omega_k}\right]
\end{split} \label{eq:revised_eom}
\end{equation}
where $ \bm{v_k}=\hbar^{-1}\partial  \e_{\bm{k}}/\partial \bm{k} $ is the group velocity, and $D\left( \bm{B},\bm{\Omega_k}\right) = \left(1+e\left( \bm{B}\cdot \bm{\Omega_k}\right)/\hbar \right)^{-1}$ is the phase space modification factor. Under the relaxation time approximation, the steady state Boltzmann equation is given by
\begin{equation}
\begin{split}
\left( \bm{\dot{r} \cdot \nabla_r} +\bm{\dot{k} \cdot \nabla_k} \right)f_{\bm{k}}=-\frac{f_{\bm{k}} - f_0}{\tau},
\end{split}      \label{eq:nonE_bte}
\end{equation}
where $\tau$ is the average scattering time between two successive collisions, $f_0 = (e^{\beta(\e_{\bm{k}}-\mu)}+1)^{-1}$ is the equilibrium Fermi-Dirac distribution function, and $f_{\bm{k}}$ is the distribution function in the presence of perturbative fields. For simplicity, in what follows we ignore the momentum and band dependence of $\tau$ and treat it to be a constant with $\tau (\mathbf{k})=\tau$.

In this paper we are interested in the NLANE of a TR invariant system where we measure a transverse electric current in response to the second-order of an applied temperature gradient $\bm{\nabla} T$ in the absence of magnetic field.  Substituting $\bf{\dot{r}}$ and $\bf{\dot{k}}$ with $\bm{E}=0$, $\bm{B}=0$ into Eq.~$\left(\ref{eq:nonE_bte}\right)$, the Boltzmann equation takes the form
\begin{equation}
\begin{split}
  \bm{v_k \cdot}\frac{\partial f_{\bm{k}}}{\partial \bm{r}} =\frac{f_0 - f_{\bm{k}} }{\tau}.
  \end{split} \label{eq:effective_bte}
\end{equation}
Now from the above equation, we can write $f_{\bm{k}}$ as
\begin{equation}
    \begin{split}
    f_{\bm{k}} &=f_0 -\tau  v_{a^{\prime}} \frac{\partial f_{\bm{k}}}{\partial r_{a^{\prime}}},
    \end{split}  \label{eq:Nernst_bte}
\end{equation}
where $a^{\prime}=x,y,z$. For consistency of notation, all the component subscripts in this section indicate the principal axes coordinates. For a system with non-zero Berry curvatures, the complete
description of the transport charge current is given by\cite{Dxiao2006_thermolelectric},
\begin{equation}
    \begin{split} \bm{j} = & -e \int \big[ d\bm{k} \big]  \bm{\dot{r}}  f_{\bm{k}}\\
    & - \bm{\nabla} \bm{\times} \frac{e}{\hbar} \int \big[d\bm{k}\big] \beta ^{-1} \bm{\Omega_k}  \operatorname{log} \left(1 + e^{-\beta \left(\e_{\bm{k}} -\mu \right) } \right)\end{split}                    \label{eq:cmlt_tc}
\end{equation}
where the first term is the usual charge current proportional to the carrier's group velocity and the second term is the intrinsic charge current supporting the transport anomaly.
From Eq.~$\left(\ref{eq:cmlt_tc}\right)$, the thermally induced charge current in the presence of a temperature gradient $ \bm{\nabla} T$ can be written as\cite{Dxiao2006_thermolelectric},
\begin{equation}
    \begin{split}
     \bm{j} = &-e\int \left[d\bm{k}\right] f_{\bm{k}} \bm{v_k} -  \frac{\bm{\nabla} T}{T}\bm{\times}\frac{e}{\hbar} \int \left[ d\bm{k} \right] \bm{\Omega _k}  \\  & \Big[ \left(\e_{\bm{k}} -\mu \right) f_{0} +\beta^{-1} \operatorname{log} \left( 1+e^{-\beta \left(\e_{\bm{k}} -\mu \right)}\right) \Big].
    \end{split}                         \label{eq:ane_current}
\end{equation}
It has been discussed in Ref.~[\onlinecite{xiaoqin2019_NLANE}] that a generalization of the above equation to the nonlinear regime could be obtained  by replacing the equilibrium distribution function $f_{0}$ by a non-equilibrium distribution function $f_{\bf{k}}$. Because of the spatial variation of the temperature, the extra terms contained in $f_{\bf{k}}$ will be proportional to the thermal gradient and therefore contribute as the NLANE response. To calculate the general expression for the NLANE coefficient, we assume the distribution function as $f_{\bm{k}}= f_0 +f_1 +f_2$, where the term $f_n$ is understood to vanish as $ \left(\bm{\nabla}  T\right)^n$ (for simplicity, $\left(\bm{\nabla}T\right)^n $ is denoted as $\bm{\nabla}^n T$ in the following). Now the Eq.~$\left( \ref{eq:Nernst_bte}\right)$ becomes
\begin{equation}
    \begin{split}
        f_1 +f_2 &=-\tau v_{a^{\prime}} \frac{\partial}{\partial r_{{a}^{\prime}}} \left( f_0 +f_1 +f_2\right),\\
        f_1 &= -\tau v_{a^{\prime}} \frac{\partial}{\partial r_{a^{\prime}}}  f_0,\\
        f_2 &=-\tau v_{a^{\prime}} \frac{\partial}{\partial r_{a^{\prime}}}f_1,
        \end{split}
\end{equation}
where $f_0$ is cancelled. After some straightforward algebra, we can obtain $f_1\left(\bm{\nabla} T \right) $, and $ f_2 \left( \bm{\nabla}^2 T\right)$ as
\begin{equation}
    \begin{split}
f_1 =& \frac{\tau}{\hbar} \left( \frac{\e_{\bm{k}}-\mu}{T} \right) \nabla_{a^{\prime}} T \frac{\partial}{\partial k_{a^{\prime}}} f_0, \\
f_2 = & -\tau ^2 v_{a^{\prime}} v_{b^{\prime}} \left(\frac{\e_{\bm{k}}-\mu}{T} \right)\frac{\partial f_0}{\partial \e_{\bm{k}} }\nabla^2_{a^{\prime} b^{\prime}} T \\
& +\frac{\tau^2}{\hbar^2}\left(\frac{\e_{\bm{k}}-\mu}{T}\right)^2 \nabla_{b^{\prime}} T \nabla_{a^{\prime}} T \frac{\partial^2 f_0}{\partial k_{a^{\prime}} \partial k_{b^{\prime}}} \\
&-\tau^2 v_{a^{\prime}} v_{b^{\prime}} \left(\frac{\e_{\bm{k}}-\mu}{T} \right) \nabla_{b^{\prime}} T \nabla_{a^{\prime}} T \left(-\frac{2}{T} \frac{\partial f_0}{\partial \e_{\bm{k}}}\right)  \\
    \end{split} \label{eq:fs}
\end{equation}
where $\nabla_{a^{\prime}} T =\partial T /\partial r_{a^{\prime}}$, is the temperature gradient along $r_{a^{\prime}}$. 
Substituting the distribution function $f_{\bm{k}} =f_0 +f_1 +f_2 $ into Eq.~$\left(\ref{eq:ane_current} \right)$, the transport charge current in response to the first-order in $\bm{\nabla} T$ can be written as
\begin{equation}
    \begin{split}
\bm{j}^0 &= - \frac{\nabla_{a^{\prime}} T}{T}\frac{e\tau}{\hbar}\int \left[d\bm{k}\right] \bm{v_{k}} \left(\e _{\bm{k}}-\mu\right)\frac{\partial f_0}{\partial k_{a^{\prime}}}
        \\
&-\frac{\bm{\nabla} T}{T}\frac{e}{\hbar} \times \int \left[ d\bm{k} \right] \bm{\Omega _k} \Big[ \left(\e_{\bm{k}} -\mu \right) f_0  +\beta^{-1} \sum_n \frac{f^n_0 }{n} \Big]
    \end{split} \label{eq:linearNernst}
\end{equation}
where the first term of the current ($\bm{j}^0$) varies linearly with the scattering time $\tau$ and is along the longitudinal direction of applied $\bm{\nabla} T$. On the other hand, the second term gives the anomalous Nernst-like current along the transverse direction of $\bm{\nabla} T$ and is independent of $\tau$.

Similarly, we can go further and calculate the non-linear Nernst-like current by extracting the terms depending on second-order in $\bm{\nabla}T$ after substituting $f_{\bm{k}}$ into Eq.~$\left(\ref{eq:ane_current} \right)$. The non-linear response of the charge current (second-order in temperature gradient) can be written as
\begin{equation}
    \begin{split}
    \bm{j} = & e\tau^2 \int \bm{v_k} \left[ d\bm{k} \right] \bigg[  v_{a^{\prime}} v_{b^{\prime}} \left(\frac{\e_{\bm{k}}-\mu}{T} \right)\left(\frac{\partial f_0}{\partial \e_{\bm{k}} }\right) \nabla^2_{a^{\prime} b^{\prime}} T \\
       & + 2 v_{a^{\prime}} v_{b^{\prime}} \left(\frac{\e_{\bm{k}}-\mu }{T^2}\right)  \left(- \frac{\partial f_0}{\partial \e_{\bm{k}}}\right) \nabla_{a^{\prime}} T \nabla_{b^{\prime}} T \\
    & -\frac{1}{\hbar^2}\left(\frac{\e_{\bm{k}}-\mu}{T}\right)^2 \frac{\partial^2 f_0}{\partial k_{a^{\prime}} \partial k_{b^{\prime}}} \nabla_{b^{\prime}} T \nabla_{a^{\prime}} T  \bigg] \\
      &- \frac{e \tau}{\hbar^2} \bm{\nabla} T \bm{\times} \int \left[ d\bm{k} \right]  \bm{\Omega _k} \frac{\partial f_0}{\partial k_{a^{\prime}}}  \left( \frac{ \e_{\bm{k}}-\mu } {T}\right)^2 \nabla_{a^{\prime}} T
    \end{split} \label{eq:nonlinear_current}
\end{equation}
where the first three terms of the above equation are purely semiclassical and Berry phase independent. In Eq.~(\ref{eq:nonlinear_current}), the contributions of the first three terms to the transport current can be distinguished from the last term based on their different orders of $\tau$ dependence, where $\tau$ is approximately picoseconds in experiments. 
Moreover, under the approximation of constant relaxation time, the first three terms will vanish in a TR invariant system because the integrand is odd under TR symmetry. 
Therefore, only the last term contributes to the non-linear Nernst-like current, which can be written as \cite{xiaoqin2019_NLANE}
\begin{equation}
\begin{split}
&\bm{j} =- \frac{e \tau}{\hbar^2} \bm{\nabla} T \bm{\times} \int \left[ d\bm{k} \right]  \nabla_{a^{\prime}} T \frac{\partial f_0}{\partial k_{a^{\prime}}} \bm{\Omega _k} \frac{\left( \e_{\bm{k}}-\mu \right)^2} {T^2}.
\end{split}             \label{eq:effc_nonlinear_current}
\end{equation}
From the above equation, the expression of non-linear Nernst current flowing in the $a$ direction can be written in a compact form as
\begin{equation}
    \begin{split}
        \bm{j}_{a^{\prime}} = \epsilon_{a^{\prime} b^{\prime} c^{\prime}}\frac{e \tau}{\hbar^2} \left(\nabla_{b^{\prime}} T\nabla_{d^{\prime}} T\right) \bm{\Lambda}_{d^{\prime} c^{\prime}} ^ T,
    \end{split} \label{eq:nlanh_current_1}
\end{equation}
where $\bm{\Lambda}_{d^{\prime} c^{\prime}} ^T$, the non-linear anomalous Nernst coefficient, is defined as
\begin{equation}
    \begin{split}
    \bm{\Lambda}_{d^{\prime} c^{\prime}} ^T = - \int \left[ d\bm{k} \right] \frac{\left( \e_{\bm{k}}-\mu \right)^2} {T^2} \frac{\partial f_0}{\partial k_{d^{\prime}}} \Omega _{\bm{k}}^{c^{\prime}}.
    \end{split}             \label{eq:lambda}
\end{equation}
Clearly, the NLANE coefficient is a pseudotensorial quantity and has a different form compared to the Berry curvature dipole which produces the NLAHE. The non-linear anomalous Nernst conductivity, which is proportional $\bm{\Lambda}_{d^{\prime} c^{\prime}}^T$, is obtained as $\frac{e \tau}{\hbar^2}\bm{\Lambda}_{d^{\prime} c^{\prime}}^T$. Therefore, unlike the linear case, this quantity is found to be a Fermi surface quantity indicating the fact that the main contribution to it comes from the states near the Fermi surface. Moreover, the NLANE coefficient is also linearly proportional to the scattering time whereas the linear anomalous Nernst coefficient is independent of $\tau$. Interestingly, the NLANE can be finite for a TR invariant system which can be easily checked by looking at Eq.~$\left(\ref{eq:lambda}\right)$.
In the presence of TR symmetry, $\bm{\Omega}_{\bm{k}}=-\bm{\Omega}_{\bm{-k}}$, $\e_{\bm{k}}=\e_{-\bm{k}}$ and $\partial f_0 /\partial k_{d^{\prime}}=-\partial f_0 /\partial \left(-k_{d^{\prime}}\right)$. Therefore, the integrand of Eq.~$\left(\ref{eq:lambda}\right)$ is an even function with respect to $k_{d^{\prime}}$ which makes the integral finite and results in a non-zero NLANE.

In three-dimension (3D), the Berry curvature $\bm{\Omega_k}$ is a pseudovector and therefore, the NLANE coefficient ($\bm{\Lambda}^T_{d^{\prime} c^{\prime}}$) becomes a pseudotensor. On the other hand, in the case of a two-dimensional (2D) system, the only nonzero component of the Berry curvature $\bm{\Omega_k}$ is $\Omega_{\bm{k}} ^{z}$ $\left(c^{\prime} =z\right)$  which is perpendicular to the $x-y$ plane indicating the fact that the Berry curvature behaves as a pseudoscalar. Therefore, in 2D the pseudotensorial quantity $\bm{\Lambda}_{d^{\prime}c^{\prime}} ^T$ is reduced to a pseudovector quantity ($\bm{\Lambda}^T_{d^{\prime}}$) confined in the 2D plane with only two independent components ($x$ and $y$ components).
Following the above discussion, Eq.~$\left( \ref{eq:effc_nonlinear_current}\right)$ can be written as
\begin{equation}
    \begin{split}
     \bm{j} = \frac{e \tau}{\hbar^2} \left( \bm{\nabla} T \cdot \bm{\Lambda}^ T\right)\bm{\nabla}T
     \bm{\times}   \bm{\hat{z}}.
    \end{split} \label{eq:nlanh_current_2}
\end{equation}
It can be shown that the largest symmetry required for a 2D system to get the non-vanishing non-linear anomalous Nernst conductivity (or $\bm{\Lambda}^ T$) is a single mirror line
(i.e. a mirror plane that is perpendicular to the 2D system). In a TR invariant 2D system, the presence of single mirror symmetry forces the $\bm{\Lambda}^ T$ to be orthogonal to the mirror plane. Moreover, according to Eq.~$\left(\ref{eq:nlanh_current_2}\right)$, when the applied temperature gradient is aligned with the direction of $\bm{\Lambda}^T$, we find that the current which flows in the transverse direction of it explicitly originate from $\bm{\Lambda}^T_{d^{\prime}}$\cite{Fu2015_NLAHE,xiaoqin2019_NLANE}.


\section{Angular dependence of Non-linear Anomalous Nernst response}\label{sec:iii}
In this section we will study the angular dependence of the NLANE for a Nernst experimental setup which is schematically shown in Fig.~\ref{fig:schemati}. Angle $\theta$ is due to the misalignment between the temperature gradient $-\bm{\nabla}T$ (red arrow) and the a-axis of the crystal (black dashed line). In Fig.~\ref{fig:schemati}(b), $M_a$ is the mirror symmetry line (black long dashed line) along $b-$axis\cite{zzDu2019_NLAHE_1,zzDu2019_NLAHE_2,kKang2019_NLAHE}. Based on the analysis in Sec. \ref{sec:ii}, $\Lambda_a$ is the only non-zero component of the NLANE coefficient that is perpendicular to the mirror line $M_a$ as shown in Fig.~\ref{fig:schemati}(b).

From Eq.~$\left(\ref{eq:linearNernst}\right)$ we could define the thermo-electric conductivity tensor $\bm{\alpha}^0$ with components as
\begin{equation}
    \begin{split}
     \alpha^0_{aa}&=-e\tau \int \left[d\bm{k}\right] v^2_a \left(\frac{\e_{\bm{k}}-\mu}{T}\right) \left(-\frac{\partial f_0}{\partial \e_{\bm{k}}}\right), \\
     \alpha^0_{ab} &=\frac{e}{\hbar} \int \left[d\bm{k}\right] {\Omega^c_{\bm{k}}} \left[\left(\frac{\e_{\bm{k}}-\mu}{T}\right) f_0 +k_B \sum_n  \frac{f_0^n}{n} \right]
    \end{split} \label{eq:linear_NC}
\end{equation}
which represent the longitudinal and transverse components of the thermo-electric  conductivity respectively\cite{Girish2016_ANE,Girish2017_ANE,Saha2018_ANE}. The linear anomalous Nernst conductivity ($\alpha^0_{ab}$) can also be written in
terms of entropy density ($s_{\bm{k}}$) as $\alpha^0_{ab}= \frac{e k_B}{\hbar} \int \left[d\bm{k}\right] \bm{\Omega_k} s_{\bm{k}}$, where  the entropy density is given by $s_{\bm{k}} =-f_0  \operatorname{log}\left(f_0\right) -(1-f_0)  \operatorname{log}\left(1-f_0\right)$. 
Now from the linear response theory, we can write
\begin{equation}
    \begin{split}
    \left(\begin{array}{cc}
        j^0 _a\\ j^0_b \end{array}\right)= \left(\begin{array}{cc}
           \alpha^0_{aa}  & \alpha^0_{ab} \\
            -\alpha^0_{ab} & \alpha^0_{bb}
        \end{array}\right)\left(\begin{array}{c}
            - \nabla_a T \\  - \nabla_b T
        \end{array}\right).
    \end{split} \label{eq:linearcMatrix}
\end{equation}
It is shown in Fig.~(\ref{fig:schemati}) that we are dealing with two sets of coordinates ($\left(x, y \right)$ and $\left(a, b \right)$) related by angle $\theta$. For this reason, we introduce the transformation matrix $\mathcal{G}(\theta)$ which transforms the principal axes $\left(a, b \right)$ coordinates to the  $\left(x, y \right)$ coordinates in experiment \cite{zzDu2019_NLAHE_1}. In the linear regime, both the charge current and thermal gradient could be transformed by $\mathcal{G}(\theta)$, like
\begin{equation}
    \begin{split}
    \left(\begin{array}{cc} j^0_a \\ j_b^0 \end{array}\right)=\mathcal{G}(\theta)  \left(\begin{array}{cc} j^0_x \\ j_y^0 \end{array}\right),
       \mathcal{G}(\theta)= \left(\begin{array}{cc} \cos{\theta} & \sin{\theta} \\ -\sin{\theta} & \cos{\theta} \end{array}\right)
    \end{split} \label{eq:transformM}
\end{equation}
Now under the transformation $\mathcal{G}(\theta)$,  we have the thermo-electric  conductivity tensor $\widetilde{\bm{\alpha}}^0 = \mathcal{G}^{\dagger} \bm{\alpha} ^0 \mathcal{G}$ given by
\begin{equation}
    \begin{split}
      \widetilde{\bm{\alpha}}^0= \left(
      \begin{array}{cc}
           \alpha^0_{bb} +\delta \alpha^0 \cos^2{\theta} &  \delta \alpha^0 \sin{\theta} \cos{\theta}+\alpha^0_{ab} \\ \delta \alpha^0 \sin{\theta} \cos{\theta}-\alpha^0_{ab}
           &  \alpha^0_{bb} +\delta \alpha^0 \sin^2{\theta}
      \end{array} \right)
    \end{split}
\end{equation}
where $\delta \alpha^0 =\alpha^0_{aa}-\alpha^0_{bb}$ is the anisotropy of the thermo-electric conductivity along the principal axes. Now, by definition the corresponding resistivity tensor $\bm{\nu}$ can be obtained as
\begin{equation}
 \begin{split}
     \left(\begin{array}{c}
            - \nabla_x T \\  - \nabla_y T
        \end{array}\right) =  \frac{1}{\widetilde{\bm{\alpha}}^0} \left(\begin{array}{cc}
        j^0 _x\\ j^0_y \end{array}\right) = \bm{\nu}\left(\begin{array}{cc}
        j^0 _x\\ j^0_y \end{array}\right),
 \end{split}    \label{eq:Nernstresistivity}
\end{equation}
with
\begin{equation}
    \bm{\nu}  =\left(
      \begin{array}{cc}
           \nu^0_{bb} +\delta \nu^0 \sin^2{\theta} & -\nu^0_{ab} -\delta \nu^0 \sin{\theta} \cos{\theta} \\ -\nu^0_{ab} +\delta \nu^0 \sin{\theta} \cos{\theta}
           &  \nu^0_{bb} +\delta \nu^0 \cos{}^2{\theta}
      \end{array} \right) \label{eq:Nernstresistivity1}
\end{equation}
where
\begin{equation}
\begin{split}
 \nu^0_{i i^{'}}=\frac{\alpha^0_{ii^{'}}}{|\widetilde{\bm{\alpha}}^0|}, \   \delta \nu^0 = \frac{\delta \alpha^0 } {|\widetilde{\bm{\alpha}}^0|} , \   |\widetilde{\bm{\alpha}}^0|= \alpha^0_{aa} \alpha^0_{bb} + (\alpha^0_{ab})^2 .
\end{split} \label{eq:Nernstresistivity2}
\end{equation}
From the Eq.~(\ref{eq:Nernstresistivity1}), the angular dependence of the thermo-electric resistivity in linear regime can be written as
\begin{equation}
\begin{split}
    \nu_{xx}= \nu^0_{bb} +\delta \nu^0 \sin^2{\theta},\\
    \nu_{xy} = -\nu^0_{ab} -\delta \nu^0 \sin{\theta} \cos{\theta}.
\end{split}    \label{eq:Nernstresistivity3}
\end{equation}

Following Eq.~$\left(\ref{eq:nlanh_current_2}\right)$, the non-linear response to the second-order of $\bm{\nabla} T$ along the principal axes can be written as,
\begin{equation}
    \left(\begin{array}{cc}  j_a \\ j_b \end{array}\right) =
    \left(\begin{array}{cccc} 0 &  \alpha_{aab} & 0 & \alpha_{abb}  \\
        \alpha_{baa} & 0 & \alpha_{bba} & 0 \end{array}\right)
        \left(\begin{array}{c}
        \nabla^2_{aa} T \\ \nabla^2_{ab} T \\ \nabla^2_{ba} T \\\nabla^2_{bb} T\end{array}                \label{eq:current tensor}
    \right)
\end{equation}
where we have a similar analogy as the non-linear Hall conductivity\cite{zzDu2019_NLAHE_1}. Here $\nabla^2_{ii^{\prime}} T =\nabla_{i} T \nabla_{i^{\prime}}T$ for $i,i^{\prime} =a,b$.
Now after the transformation by $\mathcal{G}(\theta)$, the non-linear response tensor in the experimental coordinates can be written as
\begin{equation}
\begin{split}
 \left(\begin{array}{cc} j_x \\ j_y \end{array}\right) = \bm{\widetilde{\alpha}} \left(\begin{array}{c}
        \nabla^2_{xx} T \\ \nabla^2_{xy} T \\ \nabla^2_{yx} T \\\nabla^2_{yy} T\end{array}
    \right) , \text{with   }   \bm{\widetilde{\alpha}}= \mathcal{G}^{\dagger}(\theta) \bm{\alpha}  \mathcal{G}(\theta) \otimes  \mathcal{G}(\theta)
\end{split}  \label{eq:nonlinear_current1}
\end{equation}
Therefore, the non-linear Nernst conductivity tensor $\bm{\widetilde{\alpha}}$ expressed in the experimental coordinates $\left(x,y\right)$ takes the form
\begin{equation}
    \begin{split}
     \bm{\widetilde{\alpha}} =\left(\begin{array}{cccc} 0 &  \alpha_{xxy} & 0 & \alpha_{xyy}  \\
        \alpha_{yxx} & 0 & \alpha_{yyx} & 0 \end{array}\right)
    \end{split}
\end{equation}
where all the non-zero terms have an angular dependence given as
\begin{equation}
\begin{split}
\alpha _{xxy} &=\alpha_{aab}\cos{\theta} -\alpha_{abb}\sin{\theta}, \\
\alpha_{xyy} &=\alpha_{aab} \sin{\theta} + \alpha_{abb}\cos{\theta},\\
\alpha_{yxx} & = -\alpha_{xxy},   \alpha_{yyx}=-\alpha_{xyy}
\end{split}                     \label{eq:angular depends}
\end{equation}
Similar to the NLAHE experiment, the measurable quantity in NLANE coefficient experiment is non-linear Nernst voltage. Therefore, we will now derive the angular dependence of the non-linear Nernst voltage which can be directly checked in experiments.

Based on Eq.~$\left(\ref{eq:Nernstresistivity}\right)$ and Eq.~$\left(\ref{eq:nonlinear_current}\right)$, the non-linear Nernst  voltage can be defined as
\begin{equation}
    \begin{split}
        V_x = &\nu_{xx} j_{x} +\nu_{xy} j_{y}\\
         = &\nu_{xx}\left(\alpha_{xxy} \nabla^2_{xy} T +\alpha_{xyy} \nabla^2_{yy} T\right) \\
    &  + \nu_{xy} \left(\alpha_{yxx} \nabla^2_{xx} T +\alpha_{yyx} \nabla^2_{yx} T\right).
    \end{split}
\end{equation}
Assuming the temperature gradiant along the $y$ direction and measuring the non-linear Nernst voltage along the $x$ direction (shown as the schematic setup in Fig.~(\ref{fig:schemati})), the above equation can be written with the help of Eq.~$\left(\ref{eq:Nernstresistivity}\right)$ as
\begin{equation}
    \begin{split}
      V_x =& \nu_{xx}\left(\alpha_{xxy} \nu_{xy} \nu_{yy} +\alpha_{xyy} \nu^2_{yy}\right) \left(j^0_y\right)^2 \\
      & +  \nu_{xy}\left(\alpha_{yxx} \nu^2_{xy} +\alpha_{yyx} \nu_{yy} \nu_{xy}\right) \left(j^0_y\right)^2, \\
       = & \left(\nu_{xx}\nu_{yy} -\nu^2_{xy}\right) \left(\alpha_{xxy} \nu_{xy} +\alpha_{xyy} \nu_{yy} \right) \left(j^0_y\right)^2.
    \end{split}
    \label{e5}
\end{equation}
Compare to the case of NLAHE\cite{zzDu2019_NLAHE_1}, the term $\left(\nu_{xx}\nu_{yy} -\nu^2_{xy}\right)$, which we denote as $\nu^{\prime}$, is not the same as the determinant $|\bm{\nu}| = |\widetilde{\bm{\alpha}}^0|^{-1}$. Substituting all the components in the Eq.~(\ref{e5}), the non-linear Nernst voltage takes the form
\begin{equation}
    \begin{split}
    V_x =\nu^{\prime} \left( -\alpha_{aab} \cos{\theta} +\alpha_{abb} \sin{\theta}\right)  \nu^0_{ab} \left(j^0_y\right)^2
     \\+\nu^{\prime} \left( \alpha_{aab}\sin{\theta} \nu^0_{bb} +\alpha_{abb}\cos{\theta} \nu^0_{aa} \right) \left(j^0_y\right)^2
    \end{split} \label{eq:current_dependence}
\end{equation}
Now the linear longitudinal voltage in response to the applied temperature gradiant along the $y$ direction is given by
\begin{equation}
\begin{split}
    V^0_y =\nu^0_{yy} j^0_y =\left( \nu^0_{aa}\sin^2{\theta}+ \nu ^0_{bb}\cos^2{\theta}\right) j^0_y
\end{split} \label{eq:linear_Voltage}
\end{equation}
To get rid of the current dependence ($j^0_y$) from the above expression, we take the ratio between the non-linear Nernst voltage (Eq.~(\ref{eq:current_dependence})) and linear longitudinal voltage (Eq.~(\ref{eq:linear_Voltage})) which takes the form
\begin{equation}
    \begin{split}
        \frac{V_x}{\left(V^0_y\right)^2} =\frac{\nu^{\prime} \left( -\alpha_{aab} \cos{\theta} +\alpha_{abb} \sin{\theta}\right)  \nu^0_{ab}}{\left( \nu^0_{aa}\sin^2{\theta}+ \nu ^0_{bb}\cos^2{\theta}\right)^2}
     \\+\frac{\nu^{\prime} \left( \alpha_{aab}\sin{\theta} \nu^0_{bb} +\alpha_{abb}\cos{\theta} \nu^0_{aa} \right)}{\left( \nu^0_{aa}\sin^2{\theta}+ \nu ^0_{bb}\cos^2{\theta}\right)^2}.
    \end{split}
\end{equation}

Replacing the resistivity by the conductivity, the above equation can be rewritten as
\begin{equation}
    \begin{split}
        \frac{V_x}{\left(V^0_y\right)^2} =\frac{ \nu^{\prime\prime}\left( -\alpha_{aab} \cos{\theta} +\alpha_{abb} \sin{\theta}\right)  \alpha^0_{ab}}{\left( \alpha^0_{aa}\sin^2{\theta}+ \alpha ^0_{bb}\cos^2{\theta}\right)^2}
     \\+\frac{ \nu^{\prime\prime}\left( \alpha_{aab}\sin{\theta} \alpha^0_{bb} +\alpha_{abb}\cos{\theta} \alpha^0_{aa} \right)}{\left( \alpha^0_{aa}\sin^2{\theta}+ \alpha ^0_{bb}\cos^2{\theta}\right)^2}
    \end{split} \label{eq:presimplified}
\end{equation}
where $\nu^{\prime\prime}= \alpha^0_{aa} \alpha^0_{bb}-\left(\alpha^0_{ab}\right)^2-\alpha^0_{ab}\delta \alpha^0 \sin{2\theta}$. To simplify the above equation we now define $n_0= \alpha^0_{bb}/\alpha^0_{aa}$ and $n_1 =\alpha^0_{ab}/\alpha^0_{aa}$ where $n_0$ represents the conductivity anisotropy ratio along the principal axes. Therefore, the Eq.~$\left(\ref{eq:presimplified}\right)$ takes the form
\begin{equation}
    \begin{split}
     \frac{V_x}{\left(V^0_y\right)^2} =& f_1 \alpha^0_{aa} \alpha_{aab}+f_2 \alpha^0_{aa} \alpha_{abb}
    \end{split} \label{eq:generalNernst_voltage}
\end{equation}
where the angular dependence factors $f_1$, and $f_2$ can be written in terms of $n_0,n_1$ and  $\theta$ as
\begin{equation}
\begin{split}
    f_1 =&\frac{n_0(1+n_1 \sin{2\theta})+n_1(n_1+\sin{2\theta})}{\left(\sin^2{\theta}+n_0 \cos^2{\theta}\right)^2}\left(n_0 \sin{\theta}-n_1 \cos{\theta} \right) \\
    f_2 =&\frac{n_0(1+n_1 \sin{2\theta})+n_1(n_1+\sin{2\theta})}{\left(\sin^2{\theta}+n_0 \cos^2{\theta}\right)^2}\left( n_1 \sin{\theta}+ \cos{\theta}\right)
\end{split}    \label{eq:generalNernst_angle}
\end{equation}
Eq.~$\left(\ref{eq:generalNernst_voltage}\right)$ and $\left(\ref{eq:generalNernst_angle}\right)$ give the general expressions of angular dependence of the NLANE. Moreover, for the TR invariant system the linear Nernst conductivity vanishes i.e., $n_1=0$, then we have
\begin{equation}
    \begin{split}
        f_1= \frac{n^2 _0\sin{\theta}}{\left(\sin^2{\theta}+n_0 \cos^2{\theta}\right)^2},\\
        f_2 =\frac{n _0\cos{\theta}}{\left(\sin^2{\theta}+n_0 \cos^2{\theta}\right)^2}
    \end{split} \label{eq:zero_n_1}
\end{equation}

\section{Model Hamiltonian of two-dimensional bilayer WT\MakeLowercase{e}$_2$} \label{sec:3i}
In this paper, we take the model Hamiltonian of bilayer WTe$_2$ to study the NLANE in this system. The bilayer WTe$_2$, which is created by the stacking of two monolayers\cite{kKang2019_NLAHE}, preserves the TR symmetry and contains a pair of coupled tilted Dirac nodes. Unlike the monolayer, the inversion symmetry is naturally broken in bilayer case and a tunable spin orbit coupling is allowed via electric gating\cite{Bernevig2015_WTE2,wan2011_WTE2,Shi2019_WTE2,zzDu2019_NLAHE_1}. The only crystalline symmetry that exists for bilayer WTe$_2$ is the mirror plane symmetry\cite{qMa2019_NLAHE}. Therefore, following our analysis in Sec.\ref{sec:ii}, the non-linear Nernst conductivity is expected to appear perpendicular to this mirror plane in response to an external temperature gradiant. The model Hamiltonian of bilayer WTe$_2$ can be written as
\begin{equation}
    \begin{split}
    H_{C}=\left( \begin{array}{cc}
        H_{K_1}+ P \otimes s_x & \gamma \bm{I}_0 \otimes s_x \\
         \gamma \bm{I}_0\otimes s_x & H_{K_2}+P \otimes s_x
    \end{array}\right)          \label{eq:hamiltonian_full}
    \end{split}
\end{equation}
where $P =\eta k_x \tau_z $. Here, $\eta$ is the spin-orbit coupling strength, and $\gamma$ is the hybridization strength between the two Dirac cones, whose Hamiltonian is given by,
\begin{equation}
\begin{split}
    H_{K_i}= t_x \widetilde{k}^i _x +v_0\left( k_y \sigma_x  -\zeta_i \widetilde{k}^i_x \sigma_y\right)+m \sigma_z+ E_i
\end{split}                     \label{eq:hamiltonian_spinless}
\end{equation}
where $\widetilde{k}^i_x = \left( k_x -K_i\right)$ with $K_i$ are the wave vector components of the Dirac point. Here, $t_x$ represents the tilt parameter which tilts the Dirac node along $k_x$ direction, $v_0$ is the velocity, $m$ and $E_i$ are the size of the gap and energy of the Dirac point respectively.
The $\sigma_i, s_i(i=1,2,3)$ are the Pauli matrices, and $I_0$ is a $2 \times 2$  unit matrix. $\zeta_i =\pm 1$ is the chirality of the Dirac fermions. $H_{K_i}$ with $i=1,2$ describe the individual spinless Dirac fermion from each layer. The time reversal partner of Eq.~(\ref{eq:hamiltonian_spinless}) is defined by $H
_{K_i}^{TR}(\bm{k}) =\mathcal{T}^{\dagger} H_{K_i}(\bm{-k}) \mathcal{T}$ ($\mathcal{T}= \mathcal{K}$ is the TR symmetry operator for spinless fermion, where $\mathcal{K}$ is the anti-Hermitian complex conjugation operator)\cite{Timothy2017_Weyl,Vatsal_2016_TRWeyl}, which could be written as,
\begin{equation}
    \begin{split}
      H_{K_i}^{TR}= -t_x \widetilde{k}^{i\prime} _x -v_0\left( k_y \sigma_x  +\zeta_i \widetilde{k}^{i\prime}_x \sigma_y\right)+m \sigma_z+ E_i
\end{split}                     \label{eq:hamiltonian_spinless_TR}
\end{equation}
 where $\widetilde{k}^{i\prime}_x = \left( k_x +K_i\right)$, indicating the time reversal Dirac points of the Hamiltonian in Eq.~(\ref{eq:hamiltonian_spinless}) located at $k_x = -K_{i}$. Therefore, the model Hamiltonian of bilayer WTe$_2$ given in Eq.~(\ref{eq:hamiltonian_full}) describes only half of the Brillouin zone\cite{zzDu2019_NLAHE_1,qMa2019_NLAHE}. However, studying the Hamiltonian in Eq.~(\ref{eq:hamiltonian_full}) only is sufficient to show the NLANE for bilayer WTe$_2$, since the TR partner of Eq.~(\ref{eq:hamiltonian_full}) contributes the same to NLANE.
\begin{figure}[t]
	\begin{center}
		\includegraphics[width=0.46\textwidth,height=0.44\paperheight]{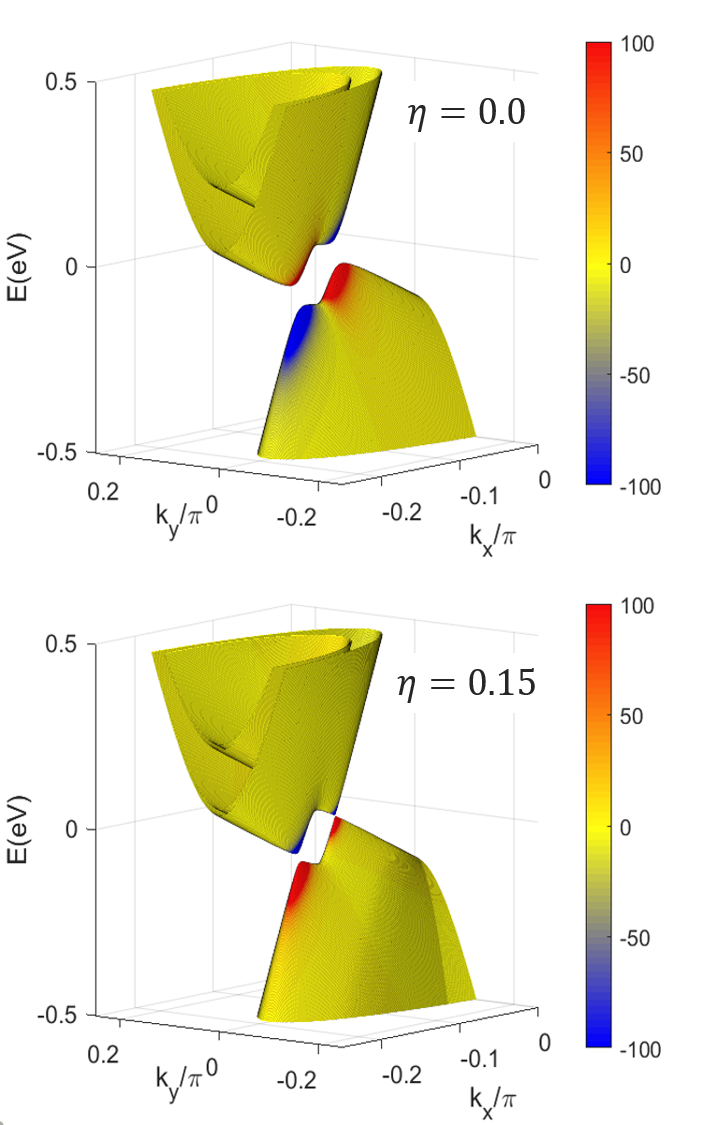}
		\llap{\parbox[b]{164mm}{\large\textbf{(a)}\\\rule{0ex}{114mm}}}
		\llap{\parbox[b]{165mm}{\large\textbf{(b)}\\\rule{0ex}{55mm}}}
	\end{center}
	\caption{(Color online) Energy dispersions of bilayer WTe$_{2}$ (a) without spin-orbit coupling ($\eta=0$) and (b) with spin-orbit coupling ($\eta= 0.15$ $eV {\angstrom}$). The colors represent the local Berry curvature distribution corresponding to each $\bm{k}$ point of the bands. A finite SOC explicitly lifts the spin degeneracy of the bands. (b) shows an anti-crossing at the band touching point with SOC of strength $\eta= 0.15$ $eV {\angstrom}$ indicating by the sign-change (change of the color) of their Berry curvatures.
	The other parameters of the Hamiltonian are $v_0=2$ $eV {\angstrom}$, $t=1.5$ $eV {\angstrom}$, $m=0.1$ $eV$, $\zeta_1 =1$, $\zeta_2=-1$, $E_1 =0.02$ $eV$, $K_1=-0.1\pi$, $K_2=-0.15\pi$, $E_2 =-0.08$ $eV$, and $\gamma=0.05$ $eV$.}   	\label{fig:Bandplot}
\end{figure}
\begin{figure}
	\begin{center}
		\includegraphics[width=0.46\textwidth,height=0.44\paperheight]{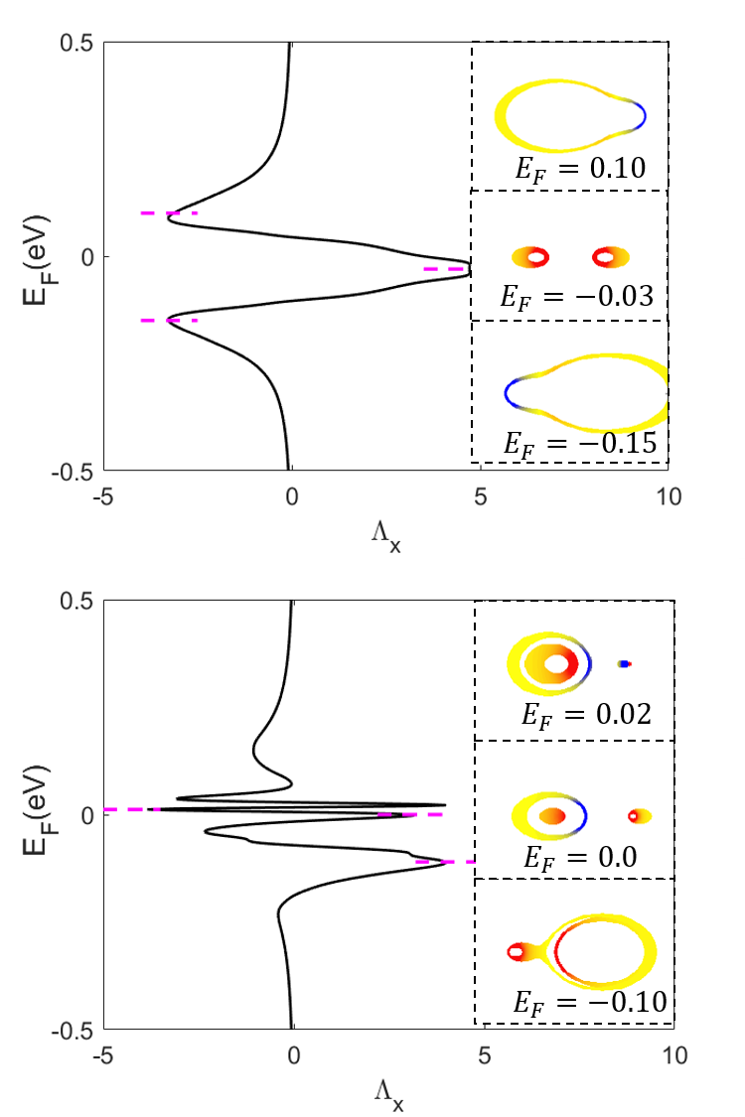}
		\llap{\parbox[b]{164mm}{\large\textbf{(a)}\\\rule{0ex}{115mm}}}
		\llap{\parbox[b]{165mm}{\large\textbf{(b)}\\\rule{0ex}{55mm}}}
	\end{center}
	\caption{(Color online) The non-linear anomalous Nernst coefficient $\Lambda_x$  as a function of chemical potential $E_F$ for different strengths of SOC (a) $\eta=0 $ and (b) $\eta=0.15$ $eV{\angstrom}$ respectively.
Insets in (a) and (b) show different Fermi surfaces where the Fermi energies are indicated by magenta dashed line. The color of the insets displays the distribution of Berry curvature weight on the Fermi surface. The color scale is the same as that in Fig.~(\ref{fig:Bandplot}). Here $T=50$ K and the other parameters used are the same as in Fig.~(\ref{fig:Bandplot}).}
	\label{fig:Nernstplot1}
\end{figure}

The energy dispersion of bilayer WTe$_{2}$ in the absence as well as in the presence of spin-orbit coupling (SOC) are shown in Fig.~$\left(\ref{fig:Bandplot}\right)$. The colors associated with each point of the bands represent the weight of the local Berry curvature. In the absence of spin-orbit coupling, the bilayer WTe$_2$, which is a semimetal with a small gap opened by the interlayer coupling, contains four tilted Dirac nodes. In Fig.~\ref{fig:Bandplot}(a) we can identify two tilted Dirac cones which carry the opposite Berry curvature because the two layers of the system are related through a mirror reflection. After turned on, the non-zero coupling $\eta$ lifts the spin degeneracy of the bands and splitted the four bands (Fig.~\ref{fig:Bandplot}(a)) into eight bands (Fig.~\ref{fig:Bandplot}(b)). With increasing the strength of the spin-orbit coupling, the band gap successively shrinks to zero and the system undergoes band inversion and then the band gap reopens again around the Dirac nodes as shown in Fig.~\ref{fig:Bandplot}(b). The Berry curvatures of the conduction and valence bands switch their sign due to the band inversion. However, with further tuning SOC, the system could become insulating\cite{Fei2017_QSHE}.

It is clearly seen from the Fig.~$\ref{fig:Bandplot}$(a) that the Berry curvature shows finite value only nearby the band edges in the absence of SOC, which is now asymmetrically distributed at $\bm{k}$ and $\bm{-k}$ at the Fermi surface (see insets of Fig.~(3)). The sign of the Berry curvature is opposite at any $\bm{k}$ point for the upper and lower bands. When we turn on the SOC, the bands split and anti-cross with their neighbors due to the hybridization factor $\gamma$. The Berry curvature of the conduction band exchanges sign with the valence band at the band inversion and band anti-crossing. In the presence of spin-orbit coupling, the Berry curvature are intensively concentrated around the Dirac point where the band anti-crossing occurs as shown in Fig.~\ref{fig:Bandplot}(b). It has been shown that around the points where the band gaps are almost vanishing, an extremely large Berry curvature originates within a small gap region i.e., a large gradiant of the Berry curvature occurs which allows a strong NLAHE in the bilayer WTe$_2$. Therefore, it is expected to get a strong NLANE response in bilayer WTe$_{2}$ based on Eq.~$\left(\ref{eq:lambda}\right)$.  We would like to point out that we can get a non-uniform distribution of the Berry curvature at a given Fermi level even without tilting in bilayer WTe$_2$ because the Dirac cones in this system are coupled due to the hybridization factor ($\gamma$) and also situated at different energies, i.e., $E_1 =0.02$eV, and $E_2=-0.08$eV.

\section{Results and Discussions} \label{sec:3ii}

Recently, non-linear anomalous Hall effect has been observed experimentally in bilayer WTe$_2$ system due to the nonuniform distribution of the Berry curvature at Fermi surface. It has been shown that the tilted band anti-crossings and band inversions lead to a large Berry curvature dipole which produces the strong non-linear Hall effect in this system. Therefore, it is expected to get a finite non-linear anomalous Nernst signal based on the Eq.~$\left(\ref{eq:lambda}\right)$ for the same system. Since the Dirac nodes are tilted along the $k_x$ direction by $t_x$ in this system, the Berry curvature has an asymmetric distribution along $k_x$ direction while symmetric distribution along $k_y$ direction as clearly seen from the figure. Therefore, combing the mirror symmetry and the TR symmetry for the bilayer WTe$_2$ system, only the $\Lambda_x^T$ is nonzero while $\Lambda_y^T =0 $. Note, the component indexes mentioned for $\bm{\Lambda}$ is within the principal axes coordinates. 

The non-linear anomalous Nernst coefficient $\Lambda_x$ $(a=x)$ at different Fermi energies in the presence as well as absence of spin-orbit coupling are depicted in Fig.~$\left( \ref{fig:Nernstplot1}\right)$.
It is clear from the figure that in the absence of SOC, the magnitude of $\Lambda_x^T$ increases and reaches to a maximum value when the chemical potential approaches towards the band edge. We would like to point out that despite of being the largest Berry curvature at the band edge, the $\Lambda_x^T$ will be zero because the group velocity vanishes at the band edge in the presence of a gap. Now, when we turned on the SOC, band inversions and band anti-crossings occur due to the hybridization factor. Therefore, sharp peaks (divergence like behavior) of the $\Lambda_x^T$ appear at the band inversions which is clearly seen from the Fig.~\ref{fig:Nernstplot1}(b). In particular, with tuning the spin-orbit coupling, the magnitude of the $\Lambda_x^T$ enhances with the shrinking gaps and the divergences appear at the band inversions.
A similar dependence of Berry curvature dipole on chemical potential has been obtained recently \cite{zzDu2019_NLAHE_1,zzDu2019_NLAHE_2}.
The insets in Fig.~$\left( \ref{fig:Nernstplot1}\right)$ show the Berry curvature distribution at the given Fermi surfaces (indicated by magenta dashed lines). We would like to mention that due to the spin-orbit coupling and hybridizations, the Berry curvature could be significantly large on the Fermi surface in bilayer system compared to the single layer WTe$_2$ system\cite{zzDu2019_NLAHE_1}.
\begin{figure}[t]
	\begin{center}
		\includegraphics[width=0.45\textwidth,height=0.44\paperheight]{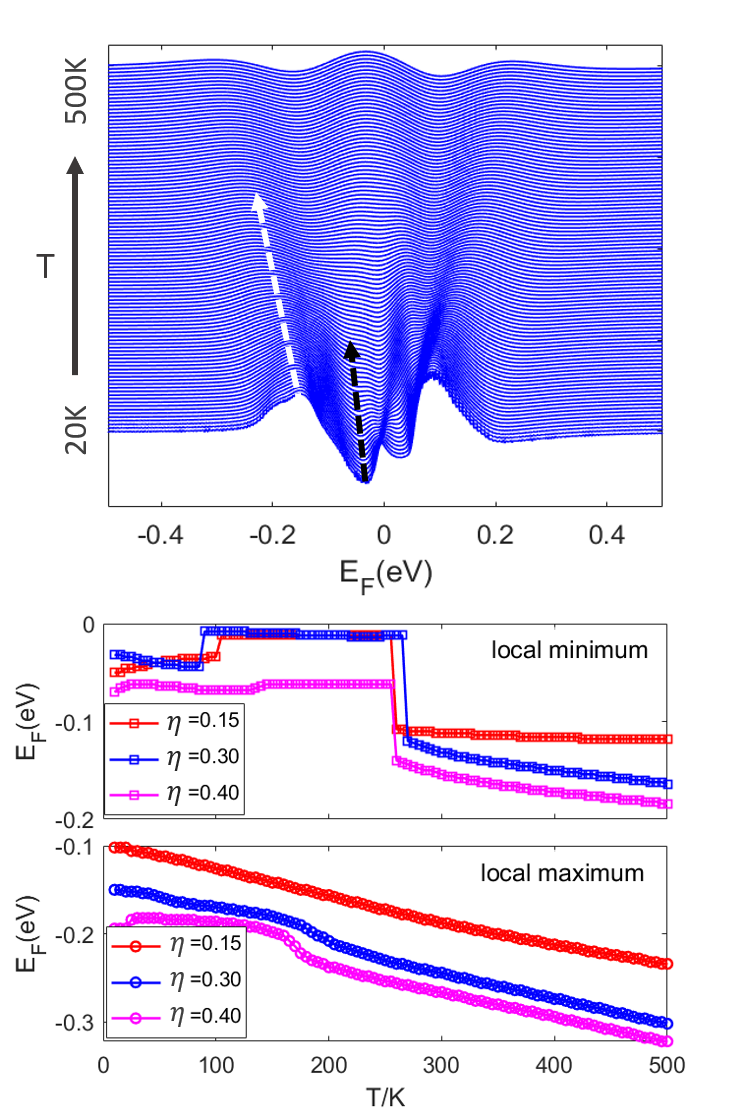}
		\llap{\parbox[b]{164mm}{\large\textbf{(a)}\\\rule{0ex}{115mm}}}
		\llap{\parbox[b]{165mm}{\large\textbf{(b)}\\\rule{0ex}{55mm}}}
	\end{center}
	\caption{(Color online) (a) Non-linear Nernst coefficient $\Lambda_x$ as a function of chemical potential ($E_F$) at different temperatures (T) for spin-orbit coupling $\eta =0.3 eV\angstrom$. The position of the local extreme points of $\Lambda_x$ shifts with increasing temperature. (b) The locus of the local minimum points (indicated by the black dashed line in panel (a)) and the locus the local maximum points (indicated by the white line in panel (a)) are shown for different spin-orbit  ($\eta$). In all cases, with increasing temperature, the chemical potential at the local minimum point shows a sudden drop while the chemical potential at the local maximum point decreases almost linearly.
}
	\label{fig:Tdependence}
\end{figure}

Now we study the temperature dependence of NLANE using the Eq~.$\left(\ref{eq:lambda}\right)$ where the temperature dependence will come from $\left(-\partial f_{\bm{k}}/\partial \e_{\bm{k}}\right)$ besides the $1/T^2$ factor. Since the Fermi-Dirac distribution function becomes broader for higher temperature, the integrand in Eq.~$\left(\ref{eq:lambda}\right)$ collects more contributions and leads to different NLANE coefficients.

The non-linear anomalous Nernst coefficient $\Lambda_x^T$  for different temperatures and chemical potentials is shown in Fig.~$\left(\ref{fig:Tdependence}a \right) $. The white arrow and the black arrow indicate the shift of the local maximum and local minimum value of $\Lambda_x^T$ in  $\left(E_F, T\right)$ parameter space. The shift of the chemical potential for these extreme points are plotted in Fig.~$\left(\ref{fig:Tdependence}b \right)$. When we increase the temperature from $20K$ to $500K$, there is a linear increasing of the magnitude of the chemical potential of the extreme points far away from the band edge ($E_F \approx 0$). On the other hand, for the extreme points nearby the band edge, there is an obvious drop in chemical potential with increasing the temperature.

This temperature-dependent phenomena could be well explained by the tilted band structure in Fig.~$\left(\ref{fig:Bandplot}\right)$. For the local maximum peak of the NLANE coefficient (bottom in Fig.~$\ref{fig:Tdependence} \left(b\right)$), its Fermi energy lies away from the band edge. The broadened Fermi distribution function $f(E_{\bm{k}},E_F, \beta)$ moves with the Fermi energy $E_F$. When the temperature increases,
the chemical potential of these extreme
points moves away from the band edge resulting in the linear increasing in negative direction of the chemical potential. On the other hand, for the local extreme points nearby the band edge (top in Fig.~$\ref{fig:Tdependence} \left( b\right)$), with an increasing temperature the chemical potential of these extreme points do not need to move to reach a minimum or maximum due to the significantly large Berry curvature around the band edge. However, the integrand for the NLANE coefficient will drop suddenly around $T=250$ K because the width of $\left(-\partial f_{\bm{k}}/\partial \e_{\bm{k}}\right)$ contains both the positive and negative Berry curvature monopoles. This explains why the chemical potential for the extreme points in Fig.~\ref{fig:Tdependence}(b) drops around $T=250$ K.

\begin{figure}[t]
	\begin{center}
		\includegraphics[width=0.46\textwidth,height=0.41\paperheight]{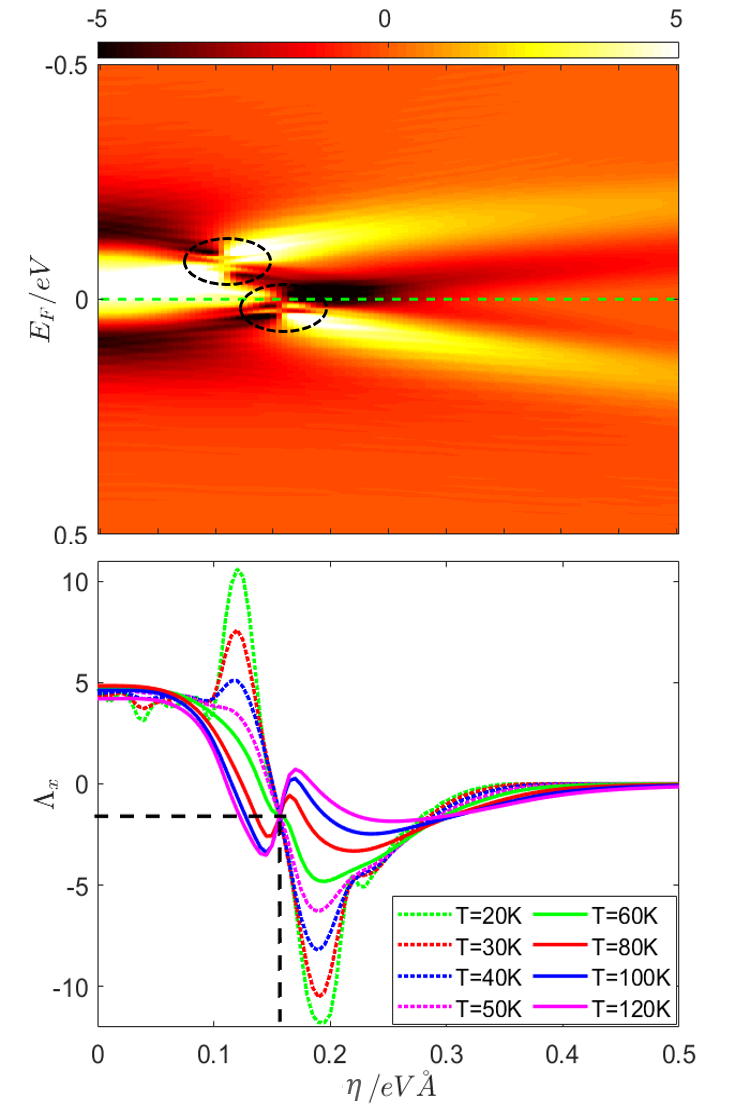}
		\llap{\parbox[b]{164mm}{\large\textbf{(a)}\\\rule{0ex}{110mm}}}
		\llap{\parbox[b]{165mm}{\large\textbf{(b)}\\\rule{0ex}{55mm}}}
	\end{center}
	\caption{(Color online)
	(a) Non-linear Nernst coefficient ($\Lambda_x^T$) at $T=50$ K with different values of the chemical potential ($E_F$) and SOC strengths ($\eta$). There are two nodes of NLANE coefficient circled by the black dashed line.
	Clearly, these two nodes behave opposite to each other 
	where this behavior of the NLANE coefficient is related to the band anti-crossing and band inversion near the Dirac nodes situated at $E_F= 0.02$~$eV$, and $E_F=-0.08$ $eV$ respectively. With increasing the SOC strength the Nernst coefficient at $E_F=0$ (indicated by the dashed line) goes from positive value (white yellow) to negative value (dark orange) showing an obvious sign change around $\eta =0.15$ $eV \angstrom$. (b) NLANE coefficient at the zero Fermi energy for different temperatures are shown.
The magnitude of the Nernst coefficient decreases with increasing the temperature. The strength of spin-orbit coupling, at which all the NLANE coefficient curves intersect, remains unaltered with changing temperature. The other parameters used here are the same as in Fig.~$\left( \ref{fig:Nernstplot1} \right)$.
	}
	\label{fig:Nernst_SO}
\end{figure}

Next we study the dependence of NLANE on the spin-orbit coupling because the SOC can affect the band structure as well as the Berry curvature distribution for the bilayer WTe$_2$ system as already discussed in the previous section. The NLANE coefficient as a function of spin-orbit coupling strength $\eta$ and chemical potential $E_F$ is depicted in Fig.~$\left(\ref{fig:Nernst_SO}\right)$. The colors indicate the magnitude of non-linear Nernst coefficient and the green dashed line is the zero energy level as shown in Fig.~$\ref{fig:Nernst_SO}\left(a\right)$. From the figure we find that there are two nodes circled by the black dashed line in the $\left(E_F, \eta\right)$ parameter space around which we can choose a local co-ordinate system. Clearly, these two nodes behave opposite to each other. In the local coordinate system, the node above $E_F=0$ carries positive conductivity (bright yellow color) in one direction whereas the other node below $E_F=0$ carries negative conductivity (dark red color) in the same direction and vice-verse in the perpendicular direction. This behavior of the NLANE coefficient is related to the band anti-crossing and band inversion near the Dirac nodes situated at $E_F= 0.02 eV$, and $E_F =-0.08 eV$ respectively in the system. Clearly, with increasing the SOC strength the non-linear Nernst coefficient for a fixed chemical potential $E_F=0$ goes from positive value to negative value showing an obvious sign change around $\eta =0.15$ $eV \angstrom$. The NLANE coefficient at $E_F=0$ for different temperatures from $T=20$ K to $T=120$ K is shown in Fig.~$\ref{fig:Nernst_SO}\left(b\right)$. It is clear from the figure that all the curves intersect around $\left(\eta = 0.15 eV \angstrom\right)$ and the strength of SOC corresponding to the intersection point remains unaltered with changing temperature whereas enhances with increasing the chemical potential ($\eta =0.12 ~eV \angstrom$ and $\eta =0.16 eV \angstrom $ for $E_F =-0.08~ eV$ and $E_F =0.02 ~eV$ respectively). The magnitude of the NLANE coefficient decreases with increasing the temperature. We find that the strength of spin-orbit coupling at which the NLANE coefficient changes sign shifts towards lower value with increasing temperature. For large value of SOC, the system becomes insulating and $E_F=0$ lies in the gap. Therefore, the NLANE coefficient at $E_F=0$ vanishes due to vanishing Fermi surface at large SOC.

\begin{figure}[t]
	\begin{center}
		\includegraphics[width=0.46\textwidth,height=0.41\paperheight]{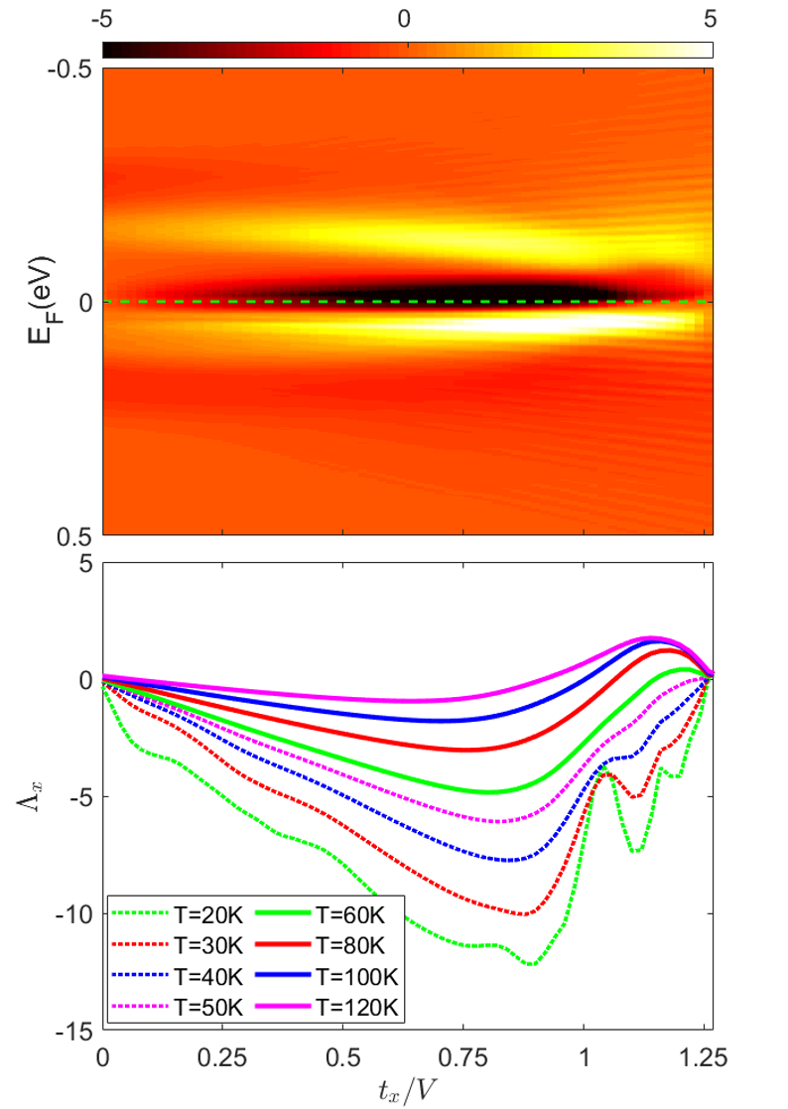}
		\llap{\parbox[b]{164mm}{\large\textbf{(a)}\\\rule{0ex}{110mm}}}
		\llap{\parbox[b]{165mm}{\large\textbf{(b)}\\\rule{0ex}{55mm}}}
	\end{center}
	\caption{(Color online) (a) Non-linear Nernst coefficient ($\Lambda_x^T$) at $T=50$ K as a function of chemical potential ($E_F$) and tilt parameter ($t_x$) in the presence of SOC $\eta =0.2 eV\angstrom$.  
Interestingly, in the absence of tilt, $\Lambda_x^T$ at the non-zero chemical potential becomes finite whereas it vanishes when $E_F$ is at zero energy. (b) $\Lambda_x^T$ at $E_F=0$ (indicated by the green dashed line in (a)) as a function of tilt parameter for different temperatures. At $t_x =1.25 v_0 $, the non-linear Nernst conductivity vanishes in bilayer WTe$_2$ at all temperatures.
The other parameters used here are the same as that in Fig.~$\left( \ref{fig:Nernstplot1} \right)$. }
	\label{fig:Nernst_tilt}
\end{figure}
\begin{figure}[t]
	\begin{center}
		\includegraphics[width=0.46\textwidth,height=0.44\paperheight]{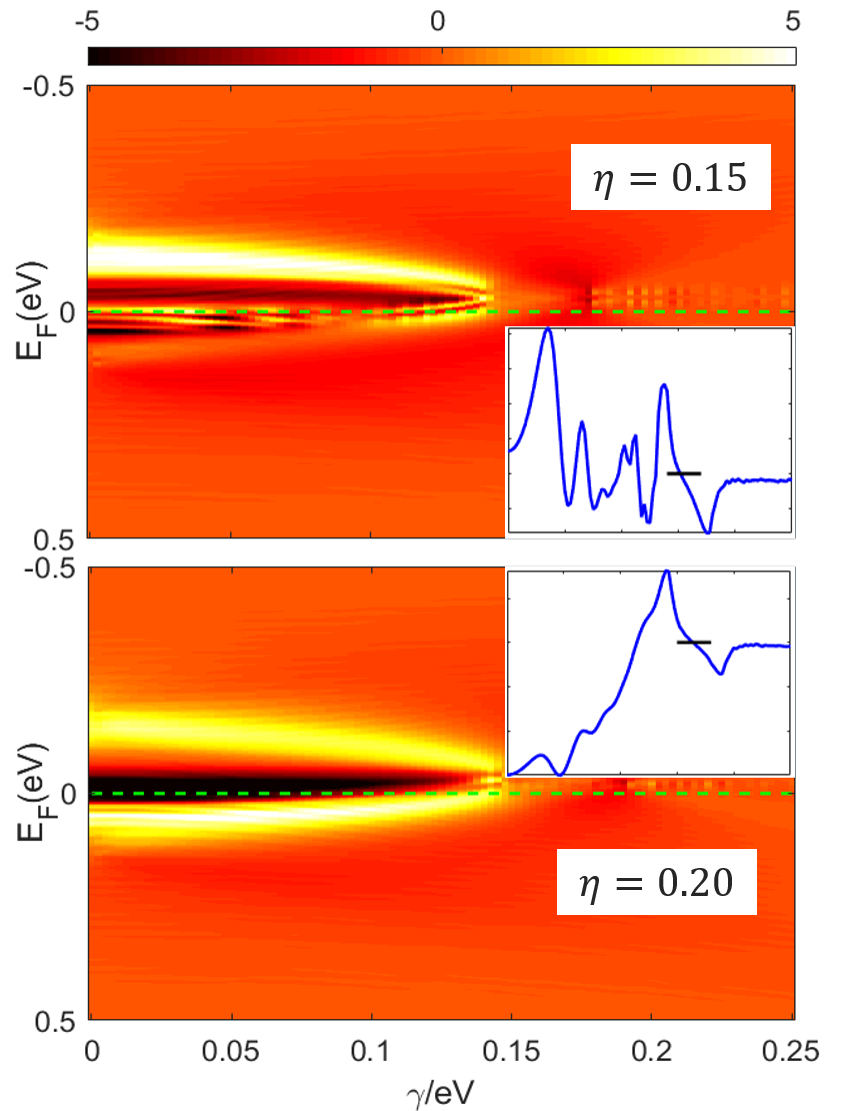}
		\llap{\parbox[b]{164mm}{\large\textbf{(a)}\\\rule{0ex}{115mm}}}
		\llap{\parbox[b]{168mm}{\large\textbf{(b)}\\\rule{0ex}{55mm}}}
	\end{center}
	\caption{(Color online) Non-linear Nernst coefficient ($\Lambda_x^T$) at $T=50$ K as a function of chemical potential ($E_F$) and inter-layer coupling ($\gamma$) in the presence of SOC $\eta =0.15 eV\angstrom$ (a) and $\eta =0.20  eV\angstrom$ (b). The magnitude of  $\Lambda_x^{T}$ decreases to zero (and stays at zero) when the inter-layer coupling goes over $\gamma=0.15 eV$, in spite of the different spin-orbit coupling in (a) and (b). The insets correspondingly show the values of $\Lambda_x^{T}$ versus $\gamma$ at $E_F=0$ (green dashed line in main figure), where the short black lines indicate the point ($\gamma\approx 0.15 eV$) above which $\Lambda_x^T=0$. All the other parameters used here are the same as in Fig.~(\ref{fig:Bandplot}).   } \label{fig:gamma_dependence}
\end{figure}
\begin{figure}[t]
	\begin{center}
		\includegraphics[width=0.46\textwidth,height=0.44\paperheight]{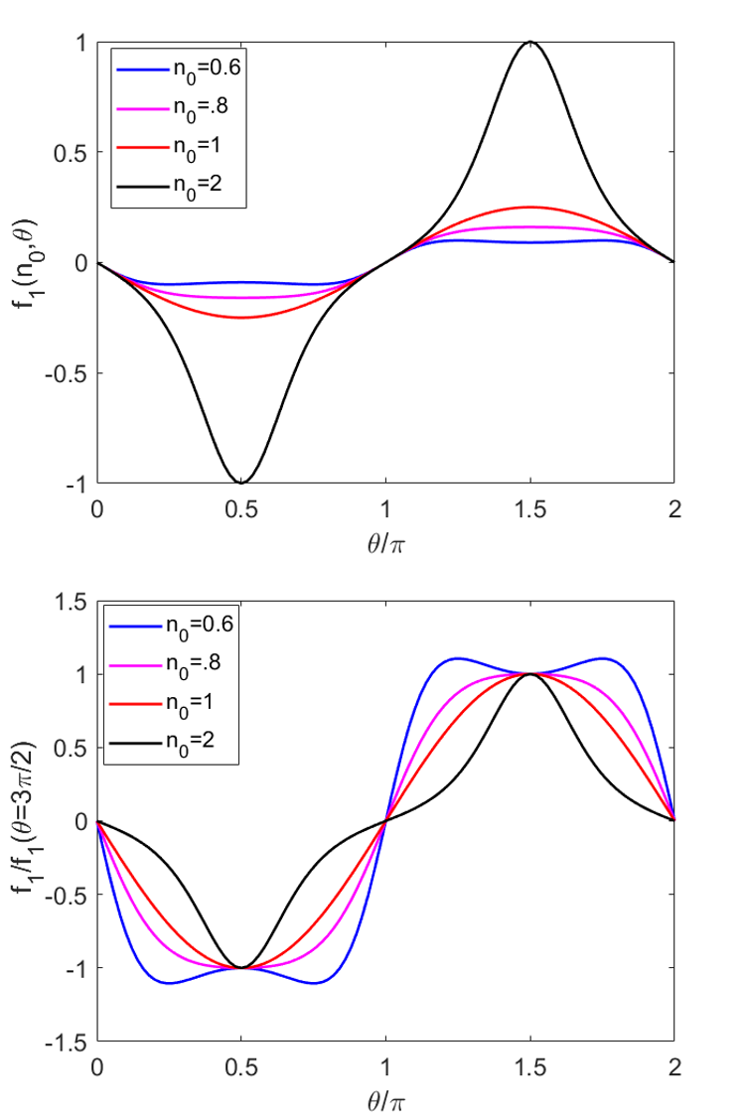}
		\llap{\parbox[b]{164mm}{\large\textbf{(a)}\\\rule{0ex}{115mm}}}
		\llap{\parbox[b]{165mm}{\large\textbf{(b)}\\\rule{0ex}{55mm}}}
	\end{center}
	\caption{(Color online) The angular dependence  factor $f_1\left(n_0, \theta \right)$ as a function of the angle $\theta$ between the temperature gradient and principal axis for different values of $n_0$.The plot in each panel (a) shows the non-scaled curves, where we see the relative magnitude of the $f_1$ for different $n_0$. While all the curves are scaled relatively to the value at $\theta =3\pi/2$ in panel (b). For a time reversal symmetric system where the linear Nernst response is zero ($n_1 =0 $), a similar angular dependence (given in Eq.~$\left(\ref{eq:zero_n_1}\right)$) of the NLANE is found as that of the NLAHE in the recent work\citep{zzDu2019_NLAHE_1}.  } \label{fig:angular_n_0}
\end{figure}
For different values of the tilt parameter, the Fermi surface gets reshaped and we have different Berry curvature distributions at the Fermi surface which result in different Nernst conductivities based on the Eq.~$\left(\ref{eq:lambda}\right)$. This fact convinces us to study the dependence of the non-linear Nernst conductivity on the tilt parameter which could be varied by tuning the intrasublattice hoppings\cite{Lucas2016_DFT}.  The $x$ component of NLANE coefficient as a function of the tile parameter ($t_x$) for different $E_F$ values is plotted in Fig.~$\ref{fig:Nernst_tilt} (a)$. For bilayer WTe$_2$ system with spin-orbit coupling $\eta = 0.2$ $eV \angstrom$, the negative non-linear Nernst conductivity arises (shown by the dark red colors) around $E_F=0$. The dependence of NLANE coefficient on the tilt parameter for different temperatures at zero chemical potential is shown in Fig.~$\ref{fig:Nernst_tilt} (b)$. In the absence of tilt, the Berry curvatures distribute symmetrically near the band edge and therefore, the NLANE coefficient vanishes. Now in the presence of tilt $t_x$, the band extrema are shifted from the original $\bm{K}(-\bm{K})$ point to opposite directions in $k_x$ axis and makes the Berry curvature distribution asymmetric at the Fermi surface along $k_x$. Therefore, as the Dirac nodes are getting tilted more and more with increasing $t_x$, the magnitude of NLANE coefficient increases and reaches a maximum around $t_x =0.8 v_0 $ (near the critical point $t_x/v_0=1$ beyond which the Dirac cone becomes overtilted). After that it decreases with further increasing $t_x$ as shown in Fig.~$\ref{fig:Nernst_tilt} (b)$. Interestingly, all the curves for different temperatures intersect around $t_x =1.25 v_0$, where the NLANE coefficient also vanishes. Moreover, in the overtilted region, we have a change in sign of the Nernst conductivity from negative to positive for high temperatures as clearly seen from the figure.
We consider both the momentum shift ($K_i$) and the energy shift ($E_i$) for the tilted Dirac fermions hybridized via a inter-layer coupling in our model (Eq.~(\ref{eq:hamiltonian_full})). Without the inter-layer coupling ($\gamma=0$), the two uncoupled Dirac fermions contribute independently, not showing any anti-crossings. A small inter-layer coupling between the two monolayers of bilayer WTe$_2$ explicitly breaks the inversion symmetry and opens a tiny gap at the band crossings of the uncoupled systems\cite{qMa2019_NLAHE,zzDu2019_NLAHE_1}. The effect of the inter-layer coupling $\gamma$ on the NLANE  $\Lambda_x^T$ for different SOC is shown in Fig.~(\ref{fig:gamma_dependence}). For both (a) and (b) $\Lambda_x^T$ decreases to zero at all Fermi levels when $\gamma$ goes over $0.15 eV$, as shown in the figure. A cross-section of the maps at $E_F=0$ (green dashed line) is shown in the inset in panel (a) and (b), where the blue line is the value for $\Lambda_x^T$  for varied $\gamma$ and the short black line indicates $\Lambda_x^T =0$ around $\gamma=0.15 eV$. This could be understood from the band structure based on the Hamiltonian model (Eq.~(\ref{eq:hamiltonian_full})) for bilayer WTe$_2$. When the coupling between the Dirac cones for bilayer WTe$_2$ increases, their Dirac points would approach each other and coalesce in momentum space despite the momentum shift $K_i$ and energy shift $E_i$. Therefore, the total NLANE for the system is reduced, due to the decrease of the unevenness of the berry curvature $\bm{\Omega}(\bm{k})$ and the Fermi distribution function $f(\bm{k})$. Increasing $\gamma$ further would open a gap for the system, rendering the whole system insulating and a zero NLANE.   

For clarity, we point out that all the components mentioned so far in this section are discussed in the principal axes coordinates, i.e., $t_x$, $\Lambda_x$ are the $a-$component (with $a=x$ is the principal axes) of the tilting and the NLANE coefficient respectively. However, the $x-$component should be distinguished from the $a-$component for the angular dependence analysis based on the setup shown in Fig.~(\ref{fig:schemati}).

In Sec.~\ref{sec:iii}, we have systematically derived the general expressions for the angular dependence of the non-linear Nernst voltage
(Eq.~$\left(\ref{eq:generalNernst_voltage}\right)$) which is tightly connected to the experiments. For a system with given parameters including
$\eta, t_x, \gamma$ and $\mu$, the angular dependence of NLANE can be shown by calculating the angular dependence factors $f_1$ and $f_2 $. In Eq.~$\left(\ref{eq:generalNernst_angle}\right)$, the conductivity anisotropy ratios $n_0$ and $n_1$ can be tuned via gate voltage in experiments. In this paper we study the angular dependence of the non-linear Nernst voltage in bilayer WTe$_2$. Considering the presence of time-reversal symmetry and the mirror plane $M_a$ of the bilayer WTe$_2$, the only non-zero component of the NLANE coefficient is $\Lambda^T_a$ which is perpendicular to the mirror plane (as shown in Fig.~(\ref{fig:schemati})). Therefore, we have $\alpha_{abb}=-\alpha_{bba} =0$ leading to the fact that we only have to calculate $f_1$ to get the full angular dependence of non-linear Nernst voltage in this system. Note, for Hamiltonian $H_C$ given in Eq.~(\ref{eq:hamiltonian_full}), both the linear AHE response ($\alpha^0_{ab}$) and the NLANE ($\alpha_{aab}$) are non-zero  based on Eq.~(\ref{eq:linearNernst}) and Eq.~(\ref{eq:effc_nonlinear_current}) respectively. Including the contribution of the TR partner of Eq.~(\ref{eq:hamiltonian_full}), the linear ANE vanishes ($n_1 =0$) while the total NLANE doubles which is indicated by its coefficient $2 \Lambda_a ^T$. 

Now we consider that the temperature gradient is applied along the $y$ direction and the non-linear Nernst voltage is measured along the $x$ direction in bilayer WTe$_2$ (shown in Fig.~(\ref{fig:schemati})). For the TR invariant system, $n_1 =0$ i.e. no linear Nernst response is present, the angular dependence of $f_1$ depends on the anisotropy ratio $n_0$ could be described by Eq.~(\ref{eq:zero_n_1}).
The angular dependence factor $f_1\left(n_0,n_1, \theta \right)$ as a function of the angle $\theta$ between the applied temperature gradient and principal axes for different values of $n_0$ with $n_1=0$ is depicted in Fig.~$\left(\ref{fig:angular_n_0} \right)$. Here, a non-scaled plot
is given in (a) where the relative magnitudes for different $n_0$ value are shown correspondingly while in (b) all the curves are scaled by the value at $\theta = 3\pi/2$. In the absence of anisotropy ($n_0=1$), the  Eq.~(\ref{eq:zero_n_1}) dictates that the angular dependence will be a $\sin \theta$ dependence which is also clearly seen from the figure. When the anisotropy ratio is large ($n_0$ $>$ 1), the angular dependence is deviated from the sine dependence. On the other hand, in the case of small anisotropy ($n_0=0.6$), a double-peak line shape appears where the maxima at $\theta =\pi/2, 3\pi/2$ turn into a minima.
All these features of the angular dependence of the non-linear Nernst voltage can be checked by tuning the anisotropy ratio through gate voltage in experiments.

\section{Conclusion} \label{sec:3iii}
In conclusion, we study the nonlinear anomalous Nernst effect of time-reversal invariant but inversion symmetry broken systems, specifically, bilayer WTe$_2$. We have  systematically derived the nonlinear Nernst current as a second order response to the temperature gradient through the Boltzmann semiclassical approach. By a symmetry analysis, we show that the transverse nonlinear Nernst response has an explicit origin in a pseudotensorial quantity, $\Lambda^T_a$, which plays a role similar to the Berry curvature dipole determining the nonlinear
anomalous Hall effect in the recent studies. We calculate and make experimental predictions for the NLANE coefficient for its dependence on temperature, spin-orbit coupling, tilting, inter-layer coupling, and chemical potential for the bilayer WTe$_2$, in which signatures of NLAHE have been observed recently. 
In addition to the Berry curvature dipole contribution, disorder mediated effects such as non-linear side-jump and skew scatterings from impurities that contribute to the non-linear anomalous Hall effect\cite{Nandy2019_NLAHE, zzDu2019_NLAHE_2,disorder_Pesin_2019,disorder_Congx_2019,Niu2019_NLAHE}, may also contribute to non-linear anomalous Nernst effect. These effects are beyond the scope of this paper and are left for future study.

Through the mapping of NLANE coefficient into the parameter space of chemical potential and spin-orbit coupling ($E_F, \eta$), we find there are two nodes that could be associated with the Dirac points of the model. Between these two nodes in the parameter space, there is a region where we find the NLANE coefficient changes sign with tuning the spin-orbit coupling.  We also show a mapping of the NLANE coefficient into the ($E_F, t_x$) and ($E_F,\gamma$) parameter space, where the tilting ($t_x$) and the inter-layer coupling ($\gamma$) effect on the NLANE could be observed respectively. We also derive the angular dependence of the NLANE in detail, where the angle is between the applied temperature gradient and the principal axes.  
Finally we wish to remark that the derivations in this work can also be applied to other $2D$ and $3D$ systems in addition to WTe$_2$. Since NLAHE has already been experimentally observed in bilayer WTe$_2$, we apply the general theoretical framework for NLANE developed in this paper to this system, and make several predictions which can be checked experimentally.




\section{Acknowledgments}
C. Z. and S. T. acknowledge support from ARO Grant
No. W911NF-16-1-0182. S. N. acknowledges MHRD, India for a research fellowship.

\bibliography{my}


\end{document}